\documentclass[12pt]{article}

\usepackage{enumitem}
\usepackage{graphicx}
\usepackage{float}
\usepackage{cite}
\usepackage{amsfonts}
\usepackage{amssymb}
\usepackage{amsmath}
\usepackage{xcolor}

\def\gtwid{\mathrel{\raise.3ex\hbox{$>$\kern-.75em\lower1ex\hbox{$\sim$}}}}
\def\ltwid{\mathrel{\raise.3ex\hbox{$<$\kern-.75em\lower1ex\hbox{$\sim$}}}}
\def\square{\kern1pt\vbox{\hrule height 1.2pt\hbox{\vrule width 1.2pt\hskip 3pt
  \vbox{\vskip 6pt}\hskip 3pt\vrule width 0.6pt}\hrule height 0.6pt}\kern1pt}

\begin{document}

\begin{titlepage}

\begin{flushright}
CCTP-2024-14 \\
ITCP/2024/14 \\
UFIFT-QG-24-06
\end{flushright}

\vskip 1cm

\begin{center}
{\bf Leading Logarithm Quantum Gravity}
\end{center}

\vskip 1cm

\begin{center}
S. P. Miao$^{1\star}$, N. C. Tsamis$^{2\dagger}$ and 
R. P. Woodard$^{3\ddagger}$
\end{center}

\vskip 0.5cm

\begin{center}
\it{$^{1}$ Department of Physics, National Cheng Kung University, \\
No. 1 University Road, Tainan City 70101, TAIWAN}
\end{center}

\begin{center}
\it{$^{2}$ Institute of Theoretical Physics \& Computational Physics, \\
Department of Physics, University of Crete, \\
GR-700 13 Heraklion, HELLAS}
\end{center}

\begin{center}
\it{$^{3}$ Department of Physics, University of Florida,\\
Gainesville, FL 32611, UNITED STATES}
\end{center}

\vspace{0cm}

\begin{center}
ABSTRACT
\end{center}
The continual production of long wavelength gravitons during primordial 
inflation endows graviton loop corrections with secular growth factors. 
During a prolonged period of inflation these factors eventually overwhelm
the small loop-counting parameter of $G H^2$, causing perturbation
theory to break down. A technique was recently developed for summing the
leading secular effects at each order in non-linear sigma models, which
possess the same kind of derivative interactions as gravity. This technique
combines a variant of Starobinsky's stochastic formalism with a variant
of the renormalization group. We generalize the new technique to quantum
gravity, resulting in a Langevin equation in which secular changes in
gravitational phenomena are driven by stochastic fluctuations of the
graviton field.

\begin{flushleft}
PACS numbers: 04.50.Kd, 95.35.+d, 98.62.-g
\end{flushleft}

\vskip 0.5cm

\begin{flushleft}
$^{\star}$ e-mail: spmiao5@mail.ncku.edu.tw \\
$^{\dagger}$ e-mail: tsamis@physics.uoc.gr \\
$^{\ddagger}$ e-mail: woodard@phys.ufl.edu
\end{flushleft}

\end{titlepage}

\section{Prologue}

The geometry of cosmology can be characterized by a 
scale factor $a(t)$ and its two first time derivatives,
the Hubble parameter $H(t)$ and the 1st slow roll 
parameter $\epsilon(t)$:
\footnote{It is more often than not convenient to employ
conformal instead of co-moving coordinates:
$ds^2 \!=\! 
-dt^2 + a^2(t) \, d{\mathbf{x}} \cdot d{\mathbf{x}} 
= 
a^2(\eta) \big[\! -d\eta^2 + a^2(\eta) \, 
d{\mathbf{x}} \cdot d{\mathbf{x}} \big]$, with
$t$ the co-moving time and $\eta$ the conformal
time.}
\begin{equation}
ds^2 = - dt^2 + a^2(t) \, d{\mathbf{x}} \cdot d{\mathbf{x}} 
\qquad \Longrightarrow \qquad 
H(t) \equiv \frac{\dot{a}}{a} 
\quad, \quad 
\epsilon(t) \equiv - \frac{\dot{H}}{H^2} 
\; . \label{geometry}
\end{equation}
In the very early universe, primordial inflation is an 
era of accelerated expansion ($H > 0$ with $0 \leq \epsilon 
< 1$). During this era virtual particles are ripped out 
of the vacuum \cite{Schrodinger:1939} and the phenomenon 
is largest for particles such as massless, minimally coupled 
(MMC) scalars and gravitons, because they are both massless 
and not conformally invariant \cite{Lifshitz:1945du,
Grishchuk:1974ny}. This particle production is thought to 
be the physical mechanism causing the primordial tensor
\cite{Starobinsky:1979ty} and scalar \cite{Mukhanov:1981xt} 
power spectra.

As inflation progresses more and more quanta are created
so that correlators which involve interacting MMC scalars 
and gravitons often show secular growth in the form of 
powers of $\ln[a(t)]$ 
\cite{Onemli:2002hr,Prokopec:2002uw,Miao:2006pn,Miao:2006gj,
Kahya:2006hc,Prokopec:2008gw,Glavan:2013jca,Wang:2014tza,
Tan:2021lza,Tan:2022xpn}. An excellent example is provided
by the theory of an MMC scalar with a quartic self-interaction
and the study of the perfect fluid form the expectation 
value of its stress tensor takes in de Sitter background 
($\epsilon = 0$):
\begin{eqnarray}
& \mbox{} & 
\mathcal{L} =  
-\frac12 \partial_{\mu} \phi \, \partial_{\nu} \phi \,
g^{\mu\nu} \sqrt{-g} - \frac{\lambda}{4!} \phi^4 \sqrt{-g} 
\; , \label{Lphi4} \\
& \mbox{} &
\langle T_{\mu\nu} \rangle = 
(\rho + p) u_{\mu} u_{\nu} + p g_{\mu\nu}
\; . \label{Tmnphi4}
\end{eqnarray}
The 2-loop dimensionally regulated and fully renormalized 
expectation value of the stress tensor equals 
\cite{Onemli:2002hr,Onemli:2004mb,Kahya:2009sz}:
\begin{eqnarray}
\rho(t) &\!\!\! = \!\!\!& 
\frac{\lambda H^4}{2^7 \pi^4} \!\times\! \ln^2(a)
+ O(\lambda^2) \; , 
\qquad \label{phi4rho} \\
p(t) &\!\!\! = \!\!\!& 
\frac{\lambda H^4}{2^7 \pi^4} \Bigl\{ 
-\ln^2(a) - \tfrac23 \ln(a) \Bigr\} + O(\lambda^2) 
\; . \qquad \label{phi4pres}
\end{eqnarray}
In the correlators of this theory for each factor of 
the coupling constant $\lambda$ up to two factors of 
$\ln(a)$ can be associated. When this bound is saturated
the contribution is known as {\it leading logarithm} 
(LLOG), for instance in the pressure (\ref{phi4pres}) 
the factor of $\, -\ln^2(a) \,$ is a leading logarithm.
Contributions which have fewer factors of $\ln(a)$ are 
known as {\it subleading logarithm}, for instance in
the pressure (\ref{phi4pres}) the factor of $\, -\frac23 
\ln(a) \,$ is subleading.

In a long period of inflation factors of $\ln[a(t)]$ can 
grow large enough to overwhelm even the smallest coupling 
constant. Obviously the most interesting particle to study 
is the carrier of the gravitational force, the graviton; 
can the universally attractive nature of the gravitational 
interaction alter cosmological parameters, kinematical 
parameters and long-range forces? A preliminary study of 
this, albeit with ``semi-primitive'' for the intended 
purpose quantum field theoretic tools, indicated a 
positive answer \cite{Tsamis:1994ca}.

After the subsequent development of the appropriate tools,
we revisit pure quantum gravity and try step by step to 
re-sum its leading logarithms and hopefully obtain the 
late time limits of cosmological correlators. The dimensionless 
coupling constant of pure quantum gravity is $G H^2$ and at 
some time the secular increase by powers of $\ln[a(t)]$ will 
overwhelm $G H^2$ causing perturbation theory to break. 
It is always a formidable affair to decipher the dynamics 
of a theory after its perturbative analysis became invalid.
While it is easy to state what is needed - a re-summation 
technique for the leading logarithms - its realization is
very hard. It is also easy to perhaps contemplate that 
developing such a technique to sum up the series of leading 
logarithms may eventually be as important for cosmology as 
the renormalization group summation of leading momentum 
logarithms was to flat space quantum field theory. The 
re-summation technique we shall consider and in our mind
has been adequately developed, is the stochastic technique 
pioneered by the late Alexei Starobinsky \cite{Starobinsky:1986fx}. 

This paper consists of five Sections and one Appendix, of which 
this Prologue was the first. In Section 2 we briefly present 
the relevant facts from pure quantum gravity in general, from 
its form in de Sitter spacetime, and summarizes the stochastic 
re-summation technique. Section 3 extends the quantum gravity 
setup to arbitrary constant $H$ backgrounds and its results 
are applied in Section 4 to obtain the desired stochastic 
(Langevin) equations that pure quantum gravity implies. 
Section 5 is the Epilogue where we discuss the physical 
implications and prospects. Finally some useful identities 
are catalogued in the Appendix.

\section{Quantum Gravity}

Pure gravity defined by the Lagrangian:
\footnote{Hellenic indices take on spacetime values
while Latin indices take on space values. Our metric
tensor $g_{\mu\nu}$ has spacelike signature
$( - \, + \, + \, +)$ and our curvature tensor equals
$R^{\alpha}_{~ \beta \mu \nu} \equiv
\Gamma^{\alpha}_{~ \nu \beta , \mu} +
\Gamma^{\alpha}_{~ \mu \rho} \,
\Gamma^{\rho}_{~ \nu \beta} -
(\mu \leftrightarrow \nu)$.}
\begin{equation}
{\mathcal L}_{inv} =
\frac{1}{\kappa^2} \big[ R \sqrt{-g} - (D-2)(D-1) H^2 \sqrt{-g} \, \big]
\; , \label{Linv}
\end{equation}
is a 2-parameter theory:
\footnote{Notice that even for a cosmological mass scale 
$\, M \!\sim\! 10^{18} GeV \,$ close to the Planck scale 
$M_{\rm Pl}$, 
the dimensionless coupling constant is very small: 
$G \Lambda \!=\! \tfrac{M^4}{M^4_{\rm Pl}} \!\sim\! 10^{-4}$.}
\begin{equation}
\kappa^2 \equiv 16 \pi G
\quad , \quad
\Lambda \equiv (D-1) H^2
\; , \label{parameters}
\end{equation}
with the following equations of motion:
\begin{equation}
R_{\mu\nu} - \frac12 R \, g_{\mu\nu} 
+ \frac12 (D-2)(D-1) H^2 g_{\mu\nu} = 0
\; . \label{eom}
\end{equation}
In terms of the full metric $g_{\mu\nu}$, the conformally
rescaled full metric ${\widetilde g}_{\mu\nu}$ and the 
graviton field $h_{\mu\nu}$ are defined thusly:
\begin{equation}
g_{\mu\nu} \equiv 
a^2 {\widetilde g}_{\mu\nu} \equiv
a^2 \big[ \eta_{\mu\nu} + \kappa h_{\mu\nu} \big] 
\; . \label{metrics}
\end{equation}

It is straightforward to express (\ref{Linv}) in terms
of the graviton field as follows:
\begin{eqnarray}
{\mathcal L}_{inv} &\!\!\!=\!\!\!&
a^{D-2} \sqrt{-{\widetilde g}} \, {\widetilde g}^{\alpha\beta}
{\widetilde g}^{\rho\sigma} {\widetilde g}^{\mu\nu} 
\Big\{ 
\tfrac12 h_{\alpha\rho, \mu} h_{\nu\sigma, \beta} 
- \tfrac12 h_{\alpha\beta, \rho} h_{\sigma\mu, \nu}
+ \tfrac14 h_{\alpha\beta, \rho} h_{\mu\nu, \sigma}
\nonumber \\
& \mbox{} & 
- \tfrac14 h_{\alpha\rho, \mu} h_{\beta\sigma, \nu}
\Big\}  
+ (\tfrac{D}{2} - 1) a^{D-1} H 
\sqrt{-{\widetilde g}} \, {\widetilde g}^{\rho\sigma}
{\widetilde g}^{\mu\nu}
h_{\rho\sigma, \mu} h_{\nu 0}
\; , \label{Linv2}
\end{eqnarray}
which is the form of $\mathcal{L}_{inv}$ we shall use
thereafter.

\subsection{The de Sitter case}

The standard paradigm of a primordial inflationary
spacetime is the de Sitter (dS) maximally symmetric
geometry:
\begin{equation}
{\widetilde g}^{dS}_{\mu\nu} = \eta_{\mu\nu}
\; . \label{dS}
\end{equation}
For our purposes we only need the Feynman rules 
associated with this geometry which allow successful
computations. Surprisingly, it seems that one 
gauge fixing choice has almost always been used to 
successfully compute Feynman loop diagrams and obtain 
dimensionally regularized and fully renormalized 
results \cite{Tsamis:1996qm,Tsamis:1996qk,Tsamis:2005je,
Miao:2005am,Kahya:2007bc,Miao:2012bj,Leonard:2013xsa,
Miao:2017vly,Glavan:2020gal,Glavan:2021adm}:
\begin{equation}
F_{\mu} =
\eta^{\rho\sigma} \, 
\big[ h_{\mu\rho, \sigma} - \tfrac12 h_{\rho\sigma, \mu} \,
- (D-2) \, a H h_{\mu\rho} \, \delta^0_{\sigma} \big]
\; . \label{FdS} \\
\end{equation}

The gauge fixing Lagrangian is the usual one:
\begin{equation}
{\mathcal L}_{GF} =
- \tfrac12 a^{D-2} \eta^{\mu\nu} \, F_{\mu} \, F_{\nu}
\; , \label{LgfdS} 
\end{equation}
and so is the ghost Lagrangian:
\begin{equation}
{\mathcal L}_{gh} =
- a^{D-2} \eta_{\mu\nu} \, {\overline c}_{\mu} \, \delta F_{\nu}
\; . \label{LghdS}
\end{equation}
In terms of the ghost and antighost fermionic fields 
$c_{\mu}$ and ${\overline c}_{\mu}$:
\begin{eqnarray}
\delta F_{\nu} 
&\!\!\! = \!\!\!& 
\eta^{\rho\sigma} \big[ \delta h_{\nu\rho, \sigma} 
- \tfrac12 \delta h_{\rho\sigma, \nu}
- (D-2) \, a H \, \delta h_{\nu\rho} \, \delta^0_{\; \sigma} \big]
\; , \label{DeltaFghdS} \\
\delta h_{\mu\nu}
&\!\!\! = \!\!\!&
- c_{\mu, \nu} - c_{\nu,\mu} - 2 a H \eta_{\mu\nu} \, c^0
   - \kappa h_{\mu\nu, \alpha} \, c^a
\; . \label{DeltahghdS}
\end{eqnarray}

In this gauge, the graviton propagator takes the form 
\cite{Tsamis:1992xa,Woodard:2004ut}:
\begin{eqnarray}
i \Bigl[ \mbox{}_{\alpha\beta} \Delta_{\rho\sigma} \Bigr](x;x') 
&\!\! = \!\!&
\Big[ 2\, {\overline \eta}_{\alpha(\rho} 
  {\overline \eta}_{\sigma)\beta}
- \tfrac{2}{D-3}\, {\overline \eta}_{\alpha\beta} 
  {\overline \eta}_{\rho\sigma} \Big] i \Delta_A (x;x')
\nonumber \\
&\mbox{}&
- 4\, \delta^0_{\; (\alpha} {\overline \eta}_{\beta)(\rho} 
  \delta^0_{\; \sigma)} \, i \Delta_B (x;x')
\nonumber \\
&\mbox{}& \hspace{-2.9cm}
+ \tfrac{2}{(D-3)(D-2)} \Big[ (D-3) \delta^0_{\; \alpha} 
  \delta^0_{\; \beta} + {\overline \eta}_{\alpha\beta} \Big]
  \Big[ (D-3) \delta^0_{\; \rho} \delta^0_{\; \sigma}
  + {\overline \eta}_{\rho\sigma} \Big] 
  i {\widetilde \Delta}_C (x;x')
\; , \qquad \label{propdS}
\end{eqnarray}
while the ghost propagator equals \cite{Tsamis:1992xa,Woodard:2004ut}:
\begin{equation}
i \Bigl[ \mbox{}_{\alpha} \Delta_{\rho} \Bigr](x;x') 
=
{\overline \eta}_{\alpha\rho} \, i \Delta_A (x;x')
- \delta^0_{\; \alpha} \delta^0_{\; \rho} \, i \Delta_B (x;x')
\; . \label{ghostpropdS}
\end{equation}

In the propagators (\ref{propdS},\ref{ghostpropdS}) 
$i \Delta_A (x;x')$ is the massless minimally coupled 
scalar propagator in de Sitter spacetime 
\cite{Onemli:2002hr,Onemli:2004mb}: 
\footnote{The de Sitter d'Alembertian operator is:
$D_A \equiv \partial_{\alpha} \big[ a^{D-2} 
\partial^{\, \alpha} \big]$.}
\begin{equation}
D_A \, i \Delta_A(x;x') = 
i \delta^D(x-x')
\; , \label{iDeltaAdS}
\end{equation}
while $i \Delta_B (x;x')$ is the massive scalar propagator
with $m^2 = (D-2)H^2$ \cite{Chernikov:1968zm}:
\begin{equation}
D_A \, i \Delta_B(x;x') = 
i \delta^D(x-x') + (D-2) H^2 a^D i \Delta_B(x;x')
\; , \label{iDeltaBdS}
\end{equation}
and $i \Delta_C (x;x')$ is the massive scalar propagator
with $m^2 = 2(D-3)H^2$ \cite{Chernikov:1968zm}:
\begin{equation}
D_A \, i \Delta_C(x;x') = 
i \delta^D(x-x') + 2(D-3) H^2 a^D i \Delta_C(x;x') 
\label{iDeltaCdS}
\end{equation}

Moreover, the graviton and ghost kinetic operators are 
respectively:
\begin{eqnarray}
D^{\mu\nu\alpha\beta} 
&\!\!\!=\!\!\!&
\big[ \tfrac12 \eta^{\mu(\alpha} \, \eta^{\beta)\nu} 
  - \tfrac14 \eta^{\mu\nu} \eta^{\alpha\beta} \big] D_A
+ (D-2) H^2 a^D \delta^{(\mu}_{\;\; 0} \, \eta^{\nu)(\alpha} 
  \delta^{\beta)}_{\;\; 0}
\; , \label{DgravdS} \\
D^{\mu\alpha}
&\!\!\!=\!\!\!&
\eta^{\mu\alpha} D_A
+ (D-2) H^2 a^D \delta^{\mu}_{\; 0} \delta^{\alpha}_{\; 0} 
\; , \label{DghostdS} 
\end{eqnarray}
and satisfy respectively:
\begin{eqnarray}
D^{\mu\nu\alpha\beta} \, 
i \Bigl[ \mbox{}_{\alpha\beta} \Delta_{\rho\sigma} \Bigr](x;x') 
&\!\!\!=\!\!\!&
\delta^{\mu}_{\; (\rho} \delta^{\nu}_{\; \sigma)} \,
i \delta^D (x-x')
\; , \label{propdScheck} \\
D^{\mu\alpha} \,
i \Bigl[ \mbox{}_{\alpha} \Delta_{\rho} \Bigr](x;x')
&\!\!\!=\!\!\!&
\delta^{\mu}_{\; \rho} \, i \delta^D (x-x')
\; . \label{ghostpropdScheck}
\end{eqnarray}

\subsection{The leading logarithm approximation}

The leading logarithm approximation (LLOG) becomes
a very essential field theoretic tool to face the 
simple fact that the presence of secular leading
logarithms eventually causes the breakdown of 
perturbation theory. As a result the only reliable
computing method has to be non-perturbative and
LLOG is such a method which seems to be appropriate 
for the specific physical environment of interest.

In this physical environment the significance of the
leading logarithms is a very slow process that requires
a very long time evolution to become noticeable due to
the smallness of the gravitational dimensionless parameter
$G\Lambda$. Hence, the graviton field $h_{\mu\nu}$ changes 
significantly less than the scale factor $a(t)$ with time;
it is {\it always} better to act derivatives on the scale 
factor $a(t)$ than to act them on $h_{\mu\nu}$.

The purpose of the LLOG technique is to sum the leading
logarithms coming from all orders of perturbation theory. 
How does this goal translate into a specific set of 
practical steps for the theories at hand? It turns out 
that after the dust settles it is only two operations 
that determine the simplified form the field equations 
take. The first of these - which we could call 
{\it ``stochastic reduction"} - operates on the classical 
level while the second - which we could call 
{\it ``integrating out"} - operates on the quantum level.

We start by considering some theory of a quantum field 
in the presence of the cosmological background 
(\ref{geometry}). Suppose the theory develops secular 
leading logarithms in its perturbative development. 
\footnote{Fields like gravitons and MMC scalars will do
precisely that.} 
The relevant question to ask is whether the interactions 
of the field possess derivatives.  
\\ [5pt]
{\bf -} The case where they do {\it not}, is the one 
Starobinsky successfully addressed in his original 
analysis of a single scalar field $\phi$ with a potential 
$V(\phi)$: 
\begin{equation}
\mathcal{L} = 
-\frac12 \partial_{\mu} \phi \, \partial_{\nu} \phi \,
g^{\mu\nu} \sqrt{-g} - V(\phi) \sqrt{-g} 
\; , \label{Lscalar}
\end{equation}
for the particular case of $V(\phi)= \tfrac{\lambda}{4!} \phi^4$ 
\cite{Starobinsky:1986fx}. Starobinsky's formalism was based 
on replacing the full field operator $\phi(t,\mathbf{x})$ with 
a stochastic field $\varphi(t,\mathbf{x})$ which commutes with 
itself $[\varphi(t,\mathbf{x}), \varphi(t',\mathbf{x}')] = 0$, 
and whose correlators are completely free of ultraviolet 
divergences. This stochastic field $\varphi(t,\mathbf{x})$ is 
constructed from the same free creation and annihilation operators 
that appear in $\phi(t,\mathbf{x})$ in such a way that the two 
fields produce the same leading logarithms at each order in 
perturbation theory. The Heisenberg field equation for $\phi$ 
gives rise to a Langevin equation for $\varphi$:
\begin{eqnarray}
\frac{\delta S[\phi]}{\delta \phi(x)}
&\!\!\! = \!\!\!& 
\partial_{\mu} \bigl[ \sqrt{-g} \, g^{\mu\nu} \partial_{\nu} \phi \bigr] 
- V'(\phi) \sqrt{-g}
\label{exact} \\
& \longrightarrow &
3 H a^3 \bigl[ \dot{\varphi} - \dot{\varphi}_0 \bigr] - V'(\varphi) a^3
\; . \label{langevin}
\end{eqnarray}
Here $\varphi_0(t,\mathbf{x})$ is a truncation of the Yang-Feldman 
free field with the ultraviolet excised and the mode function taken 
to its limiting infrared form:
\begin{equation}
\varphi_0(t,\mathbf{x}) 
\equiv \int \!\! \frac{d^3k}{(2\pi)^3} \; 
\theta\bigl( a H \!-\! k\bigr) \,
\frac{\theta(k \!-\! H) H}{\sqrt{2 k^3}} \, \Bigl\{ 
\alpha_{\mathbf{k}} \, e^{i \mathbf{k} \cdot \mathbf{x}} 
+ \alpha^{\dagger}_{\mathbf{k}} \, 
e^{-i \mathbf{k} \cdot \mathbf{x}} \Bigr\} 
\; . \label{freeYF}
\end{equation}
We can derive (\ref{langevin}) from QFT by first integrating 
the exact field equation to reach the Yang-Feldman form. We 
then note that reaching leading logarithm order requires each 
free field to contribute an infrared logarithm, so there will 
be no change to correlators, at leading logarithm order, if 
the full free field mode sum is replaced by (\ref{freeYF}). 
Differentiating this truncated Yang-Feldman equation gives 
Starobinsky's classical Langevin equation \cite{Tsamis:2005hd}. 
Furthermore, Starobinsky's technique can be proven to 
reproduce each order's leading logarithms \cite{Tsamis:2005hd} 
and - when $V(\phi)$ is bounded below - all orders can be 
summed up to give the late time limits of cosmological 
correlators in those cases for which a static limit is 
approached, as is the case for $V(\phi) = \tfrac{\lambda}{4!} 
\phi^4$ \cite{Starobinsky:1994bd}. The ``bottom line" of
this analysis is the following {\it ``stochastic reduction"} 
rule:
\\ [5pt]
${\bullet \;}$ {\it Rule for field with non-derivative 
interactions:} 
\begin{equation}
\frac{\delta S[\phi]}{\delta \phi} \Big\vert_{LLOG} 
\equiv
\frac{\delta S[\varphi]_{class}}{\delta \varphi} \Big\vert_{stoch}
= 0
\; . \label{nonDrule}
\end{equation}
The rule says that the equation of motion capturing 
the leading logarithms to all orders - the LHS of 
(\ref{nonDrule}) - is tantamount to a classical Langevin 
equation - the RHS of (\ref{nonDrule}) - derived from 
the full Heisenberg equation of motion
$\frac{\delta S[\phi]}{\delta \phi}$ by: 
\\ [3pt]
{\it (i)} At each order in the field retain only 
the terms with no derivatives and with the smallest 
number of derivatives,
\\ [3pt]
{\it (ii)} For the linear terms in the field, each 
time derivative has a stochastic source subtracted.
\\ [3pt]
Hence according to our {\it ``stochastic reduction"} 
rule, the full Heisenberg equation of motion (\ref{exact})
emanating from (\ref{Lscalar}) becomes:
\begin{eqnarray}
& \mbox{} &
\frac{\delta S[\phi]}{\delta \phi}
\;=\;
\ddot{\phi} \, + \, 3 H \dot{\phi} \, - \,
\tfrac{\nabla^2}{a^2} \, \phi + V'(\phi)
\label{exact2} \\   
& \mbox{} & \hspace{-0.9cm}
\longrightarrow \; 
\frac{\delta S[\varphi]_{class}}{\delta \varphi} 
\Big\vert_{stoch}
\!\!=\;
3 H \bigl( \, \dot{\varphi} - \dot{\varphi}_0 \, \bigr) 
+ V'(\varphi) 
\; . \label{langevin2}
\end{eqnarray}
The enormous advantage of the method becomes apparent
should we be interested in the all-orders LLOG re-summation:
we must deal with a classical stochastic equation instead 
of the Heisenberg field equations of an interacting QFT.
\\ [5pt]
{\bf -} The extension to theories with interactions that
possess derivatives was addressed in \cite{Prokopec:2007ak,
Miao:2021gic,Miao:2024atw}. We shall concentrate in theories 
with field equations containing derivative interactions of 
a single field because it is the case relevant for pure 
gravity. From the point of view of LLOG the field has a 
``dual role in the sense that: \\
{\it (i)} when undifferentiated it can and does produce
leading logarithms, \\
{\it (ii)} when differentiated it does not produce leading 
logarithms due to the action of the derivatives. \\
Thus, we need a (simple) way to isolate only the gravitons 
that contribute leading logarithms. In other words, we 
need a (simple) way which distinguishes and separates 
undifferentiated from differentiated field bilinears. 
In the presence of a constant field background the only 
bilinears that will survive are the undifferentiated ones; 
the differentiated ones contribute constants in time due 
to the action of the derivatives. Since only the 
undifferentiated bilinears furnish leading logarithms, 
we have our (simple) way at our disposal.

Therefore, in the case of a ``dual role" field the {\it 
``stochastic reduction"} rule gets augmented with the 
{\it ``integrating out"} rule which integrates out the 
differentiated field bilinears from the equations of 
motion and {\it adds} the induced result to the 
stochastically reduced equation of motion:
\\ [5pt]
${\bullet \;}$ {\it Rule for field with derivative 
interactions:}
\begin{equation}
\frac{\delta S[\Phi]}{\delta \Phi} \Big\vert_{LLOG} 
\, \equiv \,
\frac{\delta S[\varPsi]_{class}}{\delta \varPsi} \Big\vert_{stoch}
+ \, T[\varPsi] \Big\vert_{ind} \,=\, 0
\; . \label{Drule}
\end{equation}

{\bf -} Perhaps it would be appropriate, before embarking 
in quantum gravity, to review a simple and well studied  
non-linear $\sigma$-model example for a single scalar $\Phi$
\cite{Miao:2021gic}:
\begin{equation}
\mathcal{L} = 
- \big( 1 + \tfrac{\lambda}{2} \Phi \big) 
\partial_{\mu} \Phi \, \partial_{\nu} \Phi \,
g^{\mu\nu} \sqrt{-g} 
\; . \label{LPhi}
\end{equation}
A single field non-linear $\sigma$-model can be reduced 
to a free theory by a local field redefinition. Nonetheless, 
although the $S$-matrix is unity, interactions can still 
cause changes to the kinematics of $\Phi$ particles and 
to the evolution of the $\Phi$ background. 

The field equation obtained from (\ref{LPhi}) is:
\begin{equation}
\frac{\delta S[\Phi]}{\delta \Phi(x)} = 
\bigl( 1 + \tfrac{\lambda}{2} \Phi \bigr) 
\partial_{\mu} \Bigl[ \big( 1 + \tfrac{\lambda}{2} \Phi \Bigr) 
\sqrt{-g} g^{\mu\nu} \partial_{\nu} \Phi \Bigr] 
\; . \label{eomPhi}
\end{equation}

{\bf *} \underline{Dual Role I:} The {\it ``stochastic reduction"}
\\
The stochastic form of (\ref{eomPhi}) as a homogeneous 
evolution equation in co-moving time is obtained in three 
steps:
\footnote{As described above, the first step (\ref{step1}) 
follows since only time derivatives matter, the second step 
(\ref{step2}) follows since the evolution of $\Phi$ is much 
slower than that of the scale factor $a = e^{H t}$ so that 
the largest contribution comes from the external derivative 
acting on $a^3$, and the third step (\ref{step3}) follows 
from the stochastic rule whereby the full stochastic field 
$\varPsi$ has its associated stochastic jitter $\varPsi_0$ 
subtracted.}
\begin{eqnarray}
\frac{\delta S[\Phi]}{\delta \Phi(x)} 
&\!\! \longrightarrow \!\!&
-\bigl( 1 + \tfrac{\lambda}{2} \Phi \bigr) 
\frac{d}{dt} \Bigl[ \big( 1 + \tfrac{\lambda}{2} \Phi \bigr) 
a^3 \dot{\Phi} \Bigr]
\label{step1} \\
&\!\! \longrightarrow \!\!&
- 3H a^3 \bigl( 1 + \tfrac{\lambda}{2} \Phi \bigr)^2 \dot{\Phi}
\label{step2} \\
&\!\! \longrightarrow \!\!&
- 3H a^3 \bigl( 1 + \tfrac{\lambda}{2} \varPsi \bigr)^2 
\big[ \dot{\varPsi} - \dot{\varPsi}_0 \big]
\equiv \,
\frac{\delta S[\varPsi]_{class}}{\delta \varPsi} \Big\vert_{stoch}
\; . \label{step3}
\end{eqnarray}

{\bf *} {\underline{Dual Role II}:} The {\it ``integrating out"}
\\
We must add to the stochastic equation of motion 
(\ref{step3}) the induced effective force arising 
from the contribution of undifferentiated fields 
$\Phi$ in the constant background $\Phi_0$:
\footnote{Again only time derivatives matter, while 
$\partial_{\nu} \langle \Phi^2 \rangle \vert_{\Phi_0} 
= 
2 \delta^0_{\; \nu} H \tfrac{H^2}{8\pi^2}
\big( 1 + \frac{\lambda}{2} \Phi_0 \big)^{\! -2}$.
Moreover the process of integrating out, for instance,
singly or doubly differentiated scalar bilinears amounts
to replacing them with singly or doubly differentiated
scalar propagators in the presence of a spacetime 
constant scalar, and this is tantamount to changing 
the scalar field strength \cite{Miao:2021gic}.}
\begin{eqnarray}
\frac{\delta S[\Phi]}{\delta \Phi(x)} 
&\!\! \longrightarrow \!\!&
- \bigl( 1 + \tfrac{\lambda}{2} \Phi_0 \bigr) 
\frac{d}{dt} \Bigl[ \tfrac{\lambda}{4} a^3 
\tfrac{d}{dt} \langle \Phi^2 \rangle_{\Phi_0} \Bigr]
\label{Step1} \\
&\!\! \longrightarrow \!\!&
- \big( 1 + \frac{\lambda}{2} \Phi_0 \big)
\frac{d}{dt} \Big[ \tfrac{\lambda}{4} a^3 \tfrac{H^2}{4\pi^2} 
   \bigl( 1 + \tfrac{\lambda}{2} \Phi_0 \bigr)^{-2} \Big]
\label{Step2} \\
&\!\! \longrightarrow \!\!&
- \frac{3 \lambda H^4}{16\pi^2} \,
\frac{a^3}{\bigl( 1 + \tfrac{\lambda}{2} \varPsi \bigr)}
\equiv \, T[\varPsi] \Big\vert_{ind}
\; , \label{Step3}
\end{eqnarray}
where we have replaced the constant field $\Phi_0$ with
the spacetime field $\varPsi$.

Adding the two contributions (\ref{step3},\ref{Step3})
gives the desired Langevin equation associated with 
(\ref{LPhi}):
\begin{eqnarray}
\frac{\delta S[\Phi]}{\delta \Phi} \Big\vert_{LLOG} 
&\!\!\! \equiv \!\!\!&
\frac{\delta S[\varPsi]_{class}}{\delta \varPsi} \Big\vert_{stoch}
+\, T[\varPsi] \Big\vert_{ind} 
\nonumber \\
&\!\!\! = \!\!\!& 
- 3H a^3 \bigl( 1 + \tfrac{\lambda}{2} \varPsi \bigr)^2 
  \big[ \dot{\varPsi} - \dot{\varPsi}_0 \big]
- \frac{3 \lambda H^4}{16\pi^2} \,
  \frac{a^3}{\bigl( 1 + \tfrac{\lambda}{2} \varPsi \bigr)}
= 0
\qquad\qquad \label{dualrule} \\
\Longrightarrow \;
\dot{\varPsi} 
&\!\!\! = \!\!\!&
\dot{\varPsi_0}
- \frac{\lambda H^3}{16\pi^2} \,
\frac{1}{\bigl( 1 + \tfrac{\lambda}{2} \varPsi \bigr)^3}
\; . \label{langevindual}
\end{eqnarray}
We should mention that the above procedure has been 
thoroughly checked against perturbative computations
up to 2-loop order and the highly non-trivial agreement 
is complete \cite{Miao:2021gic}.
\\ [5pt]
{\bf -} We conclude by noting that to arrive at the 
elusive equations which describe LLOG pure quantum 
gravity, we simply have to effect the two operations 
which will allow us to do that:
\footnote{Although only theories with a single field 
were discussed, it is clear that the same operations 
apply to theories with many fields. Such analysis can 
be more intricate when some of the fields can produce 
leading logarithms - e.g. gravitons, MMC scalars - while 
others cannot - e.g. fermions, photons, conformally 
coupled scalars. Examples of such theories which were 
fully studied can be found in \cite{Prokopec:2007ak,
Miao:2021gic,Miao:2024atw}.}
\\
{\it (i)} the {\it ``stochastic reduction"} of the 
field equations to a classical Langevin equation, \\
{\it (ii)} the {\it ``integrating out"} of the 
differentiated fields in a constant background
to obtain the induced stress tensor.

\section{The extension to any constant graviton \\
background}

In order to accomodate the LLOG approximation in pure 
gravity (\ref{Linv}) we should like to integrate out
the differentiated graviton fields in the presence of
a constant graviton background. Thus, we first consider 
the general class of conformally rescaled backgrounds 
with constant $H$ and arbitrary $\widetilde{g}_{\mu\nu}$ 
because the ultimate object of our study is afterall 
the time evolution of constant $H$ spacetimes:
\begin{equation}
g_{\mu\nu}(x) \equiv 
a^2 \, \widetilde{g}_{\mu\nu}(x) \equiv 
a^2 \big[ \eta_{\mu\nu} + \kappa h_{\mu\nu}(x) \big] 
\quad , \quad 
a = -(H \eta)^{-1}
\; . \label{background}
\end{equation}
When we restrict for LLOG reasons  to constant 
$\widetilde{g}_{\mu\nu}$ the curvature tensor 
takes the form:
\begin{equation}
R^{\rho}_{~\sigma\mu\nu} \Big\vert_{\widetilde{g}_{\mu\nu} = c}
=
-H^2 \, \widetilde{g}^{00} \bigl(
\delta^{\rho}_{~\mu} \, g_{\sigma \nu}
- \delta^{\rho}_{~\nu} \, g_{\sigma\mu} \bigr)
\; , \label{RiemannConstant}
\end{equation}
and we recognize a de Sitter geometry albeit with 
a different cosmological constant:
\footnote{The process of integrating out, for instance, 
singly or doubly differentiated graviton bilinears 
amounts to replacing them with singly or doubly 
differentiated graviton propagators in the presence 
of a constant graviton background, and this is 
tantamount to replacing them with singly or doubly 
differentiated de Sitter graviton propagators with 
a different Hubble parameter.}
\begin{equation}
\widetilde{g}_{\mu\nu , \rho} = 0
\quad \Longrightarrow \quad
H^2
\; \longrightarrow \;
-\widetilde{g}^{00} H^2
\; . \label{constantHshift}
\end{equation}

However, because gravitons have tensor indices, the gauge 
fixing procedure must be extended to accomodate the arbitrary 
constant graviton background. The most efficient way to 
extend the Feynman rules from de Sitter to any spacetime 
such that $\, {\widetilde g}_{\mu\nu, \rho} = \,0$, starts 
with the 3+1 decomposition.

\subsection{The 3+1 decomposition}

The standard 3+1 decomposition of the metric tensor
was pioneered by Arnowit, Deser, and Misner (ADM) 
to formulate the Hamiltonian dynamics of general
relativity \cite{Arnowitt:1962hi}. The full metric 
is expressed in terms of the lapse function $N$, 
the shift function $N^i$, and the spatial metric
$\gamma_{ij}$:
\begin{eqnarray}
{\widetilde g}_{\mu\nu} 
&\!\!\!\! = \!\!\!\!&
\begin{pmatrix}
-N^2 \!+\! \gamma_{kl} N^k N^l \;&\; -\gamma_{jl} N^l \\
\\
-\gamma_{ik} N^k & \gamma_{ij} \\
\end{pmatrix}
\label{3+1lower} \\
&\!\!\!\! = \!\!\!\!&
\begin{pmatrix}
\gamma_{kl} N^k N^l \;&\; -\gamma_{jl} N^l \!&\! \\
\\
-\gamma_{ik} N^k \;&\; \gamma^{ij} \\
\end{pmatrix}
-
\begin{pmatrix}
-N \\
\\
0 \\
\end{pmatrix}
_{\!\!\!\mu}
\begin{pmatrix}
-N \\
\\
0 \\
\end{pmatrix}
_{\!\!\!\nu} 
\equiv
{\overline \gamma}_{\mu\nu} \!\!- u_{\mu} u_{\nu}
\; , \qquad \label{3+1lowerB}
\end{eqnarray}
which implies the following form for its inverse: 
\begin{eqnarray}
{\widetilde g}^{\mu\nu}
&\!\!\!\! = \!\!\!\!&
\begin{pmatrix}
-\frac{1}{N^2} & -\frac{N^j}{N^2} \\
\\
-\frac{N^i}{N^2} \;&\; \gamma^{ij} \!-\! \frac{N^i N^j}{N^2} \\
\end{pmatrix}
\label{3+1upper} \\
&\!\!\!\! = \!\!\!\!&
\begin{pmatrix}
0 & 0 \\
\\
0 \;&\; \gamma^{ij} \\
\end{pmatrix}
-
\begin{pmatrix}
\frac{1}{N} \\
\\
\frac{N^i}{N} \\
\end{pmatrix}
^{\!\!\!\mu}
\begin{pmatrix}
\frac{1}{N} \\
\\
\frac{N^j}{N} \\
\end{pmatrix}
^{\!\!\!\nu} 
\equiv
{\overline \gamma}^{\mu\nu} \!\!- u^{\mu} u^{\nu}
\; . \qquad \label{3+1upperB}
\end{eqnarray}
In relations (\ref{3+1lowerB},\ref{3+1upperB}) we have 
expressed the 3+1 decomposition in the form convenient 
for our purposes; ${\overline \gamma}^{\mu\nu}$ is the 
``spatial part" and $u^{\mu}$ is the ``temporal part".

Finally, in the Appendix some identities associated
with the 3+1 decomposition just described are recorded
(\ref{Htilde},\ref{u's}).

\subsection{The Gauge Fixing Extension}

In analogy with (\ref{LgfdS}), the extended gauge fixing 
Lagrangian term equals:
\begin{equation}
{\widetilde{\mathcal L}}_{GF} =
- \tfrac12 a^{D-2} \sqrt{-{\widetilde g}} \, 
{\widetilde g}^{\mu\nu} \, {\widetilde F}_{\mu} \,
{\widetilde F}_{\nu}
\; , \label{Lgf1} \\
\end{equation}
and, similarly, in analogy with (\ref{FdS}) and taking 
into account (\ref{u's}), the extended gauge condition 
becomes:
\begin{equation}
{\widetilde F}_{\mu} =
{\widetilde g}^{\rho\sigma} \, 
\Big[ h_{\mu\rho, \sigma} - \tfrac12 h_{\rho\sigma, \mu} \,
- (D-2) a {\widetilde H} h_{\mu\rho} u_{\sigma} \Big]
\quad , \quad 
{\widetilde H} \equiv \frac{H}{N}
\; . \qquad \label{F} 
\end{equation}

Substituting (\ref{F}) into (\ref{Lgf1}) we arrive at 
the desired form for the gauge fixing Lagrangian term:
\begin{eqnarray}
{\widetilde{\mathcal L}}_{GF}
&\!\!\!=\!\!\!&
a^{D-2} \sqrt{-{\widetilde g}} \, {\widetilde g}^{\alpha\beta}
{\widetilde g}^{\rho\sigma} {\widetilde g}^{\mu\nu} 
\Big\{ 
- \tfrac12 h_{\mu\rho, \sigma} h_{\nu\alpha, \beta} 
+ \tfrac12 h_{\mu\rho, \sigma} h_{\alpha\beta, \nu}
- \tfrac18 h_{\rho\sigma, \mu} h_{\alpha\beta, \nu}
\qquad \nonumber \\
& \mbox{} & 
+ (D-2) a {\widetilde H} h_{\mu\rho, \sigma} h_{\nu\alpha} u_{\beta}
- (\tfrac{D}{2}-1) a {\widetilde H} h_{\rho\sigma, \mu} h_{\nu\alpha} 
u_{\beta}
\nonumber \\
& \mbox{} & 
- \tfrac12 (D-2)^2 a^2 {\widetilde H}^2 h_{\mu\rho} h_{\nu\alpha} 
u_{\beta} u_{\sigma}
\Big\} 
\label{Lgf2}
\end{eqnarray}

\subsection{The Graviton Propagator}

In analogy with (\ref{propdS}), and using the 3+1
decomposition (\ref{3+1lowerB}) and identities (\ref{u's}), 
we deduce the extension for the graviton propagator:
\begin{eqnarray}
i \Bigl[ \mbox{}_{\alpha\beta} 
  {\widetilde \Delta}_{\rho\sigma} \Bigr](x;x') 
&\!\! = \!\!&
\Big[ 2\, {\overline \gamma}_{\alpha(\rho} 
  {\overline \gamma}_{\sigma)\beta}
- \tfrac{2}{D-3}\, {\overline \gamma}_{\alpha\beta} 
  {\overline \gamma}_{\rho\sigma} \Big]
  i {\widetilde \Delta}_A (x;x')
\label{prop} \\
&\mbox{}&
- 4\, u_{(\alpha} {\overline \gamma}_{\beta)(\rho} u_{\sigma)}
  \, i {\widetilde \Delta}_B (x;x')
\nonumber \\
&\mbox{}& \hspace{-2.9cm}
+ \tfrac{2}{(D-3)(D-2)} \Big[ (D-3) u_{\alpha} u_{\beta} 
  + {\overline \gamma}_{\alpha\beta} \Big]
  \Big[ (D-3) u_{\rho} u_{\sigma}
  + {\overline \gamma}_{\rho\sigma} \Big] 
  i {\widetilde \Delta}_C (x;x')
\; . \nonumber
\end{eqnarray}

The corresponding quadratic operator equals: 
\begin{equation}
\widetilde{\mathcal D}^{\mu\nu\alpha\beta}
=
\tfrac12 \Big[ {\widetilde g}^{\mu(\alpha} \, {\widetilde g}^{\beta)\nu}
  - \tfrac12 {\widetilde g}^{\mu\nu} \, {\widetilde g}^{\alpha\beta} \Big] 
  \widetilde{\mathcal D}_A
+ (D-2){\widetilde H}^2 a^D \sqrt{- {\widetilde g}} \,
  u^{(\mu} \, {\widetilde g}^{\nu)(\alpha} \, u^{\beta)}
\; , \label{Dgrav}
\end{equation}
and allows us to check that indeed the proper condition
is satisfied:
\begin{equation}
\widetilde{\mathcal D}^{\mu\nu\alpha\beta} \, 
i \Bigl[ \mbox{}_{\alpha\beta} \widetilde{\Delta}_{\rho\sigma} \Bigr](x;x') 
=
\delta^{\mu}_{\; (\rho} \delta^{\nu}_{\; \sigma)} \,
i \delta^D (x-x')
\; . \label{propcheck}
\end{equation}

Finally, in the Appendix we have catalogued the coincidence
propagator limits (\ref{proplimA+B}-\ref{ddproplimC}) of
potential interest and we should like to emphasize that of
these only $\, i {\widetilde \Delta}_A (x;x') \,$ contains 
a divergence.

\subsection{The Ghost contribution}
The contribution to the action of the ghost and antighost
fermionic fields $c_{\mu}$ and ${\overline c}_{\mu}$ is 
the usual one:
\begin{equation}
{\mathcal L}_{gh} =
- a^{D-2} \sqrt{-{\widetilde g}} \, {\widetilde g}^{\mu\nu}
{\overline c}_{\mu} \, \delta F_{\nu}
\; , \label{Lgh}
\end{equation}
where in the infinitesimal variation $\delta F_{\nu}$ of the
gauge fixing functional (\ref{F}) the variation parameter is 
the ghost field $c$:
\begin{eqnarray}
\delta F_{\nu} 
&\!\!\!=\!\!\!& 
{\widetilde g}^{\rho\sigma} \Big[ 
\delta h_{\nu\rho, \sigma} - \tfrac12 \delta h_{\rho\sigma, \nu}
- (D-2) a {\widetilde H} \, \delta h_{\nu\rho} \, u_{\sigma} \Big]
\nonumber \\
& \mbox{} &
+ \delta {\widetilde g}^{\rho\sigma} \Big[ 
h_{\nu\rho, \sigma} - \tfrac12 h_{\rho\sigma, \nu}
- (D-2) a {\widetilde H} \, h_{\nu\rho} \, u_{\sigma} \Big]
\; , \label{deltaFgh}
\end{eqnarray}
so that:
\begin{eqnarray}
\delta h_{\mu\nu, \rho} 
&\!\!\!=\!\!\!&
- c^{\alpha}_{\; ,\mu} \, {\widetilde g}_{\alpha\nu}
- c^{\alpha}_{\; ,\nu} \, {\widetilde g}_{\alpha\mu}
+ 2 a {\widetilde H} \, {\widetilde g}_{\mu\nu} u_{\alpha} c^{\alpha}
- \kappa h_{\mu\nu, \alpha} \, c^{\alpha}
\label{deltahgh} \\
\delta {\widetilde g}^{\rho\sigma} 
&\!\!\!=\!\!\!&
- {\widetilde g}^{\rho\alpha} {\widetilde g}^{\sigma\beta}
\, \kappa \, \delta h_{\alpha\beta}
\; . \label{deltag}
\end{eqnarray}

The ghost propagator (\ref{ghostpropdS}) generalizes to:
\begin{equation}
i \Bigl[ \mbox{}_{\alpha} \Delta_{\rho} \Bigr](x;x') 
=
{\overline \gamma}_{\alpha\rho} \, i {\widetilde \Delta}_A (x;x')
- u_{\alpha} u_{\rho} \, i {\widetilde \Delta}_B (x;x')
\; . \label{ghostprop}
\end{equation}
The action of the ghost kinetic operator:
\begin{equation}
D^{\mu\alpha}
=
{\widetilde g}^{\mu\alpha} {\widetilde D}_A 
+ (D-2) {\widetilde H}^2 a^D {\sqrt {-\widetilde g}} \,
u^{\mu} u^{\alpha} 
\; , \label{ghostquadratic} 
\end{equation}
on (\ref{ghostprop}) gives:
\begin{equation}
\widetilde{\mathcal D}^{\mu\alpha} \, 
i \Bigl[ \mbox{}_{\alpha} \widetilde{\Delta}_{\rho} \Bigr](x;x') 
=
\delta^{\mu}_{\; \rho} \, i \delta^D (x-x')
\; . \label{ghostpropcheck}
\end{equation}

Substituting (\ref{deltahgh},\ref{deltag}) in (\ref{deltaFgh})
and the resulting expression in (\ref{Lgh}), we obtain after
some cumbersome but straightforward tensor algebra the following 
expression for the ghost Lagrangian:
\begin{eqnarray}
{\mathcal L}_{gh}
&\!\!\!=\!\!\!&
\sqrt{-{\widetilde g}} \, {\widetilde g}^{\rho\sigma}
  {\overline c}_{\alpha} \partial_{\rho} 
  ( a^{D-2} \partial_{\sigma} c^{\alpha} )
+ (D-2) a^D {\widetilde H}^2 \sqrt{-{\widetilde g}} \, 
  {\widetilde g}^{\alpha\beta} {\overline c}_{\alpha}
  u_{\beta} u_{\rho} \, c^{\rho}
\nonumber \\
& \mbox{} & \hspace{-0.9cm}
- a^{D-2} \, {\widetilde g}^{\alpha\beta} {\widetilde g}_{\gamma\rho} \,
  \partial_{\sigma}( \sqrt{-{\widetilde g}} \, {\widetilde g}^{\rho\sigma} )
  \, c^{\gamma}_{\; ,\beta} \, {\overline c}_{\alpha}
\nonumber \\
& \mbox{} & \hspace{-0.9cm}
+ a^{D-2} \sqrt{-{\widetilde g}} \, {\widetilde g}^{\alpha\beta}
  {\overline c}_{\alpha} c^{\gamma} \, \partial_{\gamma}
  ( {\widetilde g}^{\rho\sigma} \kappa h_{\beta\rho, \sigma}
  - \tfrac12 {\widetilde g}^{\rho\sigma} \kappa h_{\rho\sigma, \beta} )
\nonumber \\
& \mbox{} & \hspace{-0.9cm}
+ a^{D-1} {\widetilde H} \sqrt{-{\widetilde g}} \, {\overline c}_{\alpha}
  \big\{ (D-2) {\widetilde g}^{\alpha\beta}_{\;\;\; ,\gamma} 
  u_{\beta} \, c^{\gamma}
+ (D-2) {\widetilde g}^{\alpha\beta} \, {\widetilde g}^{\rho\sigma}
  \, \kappa h_{\beta\rho} u_{\gamma} c^{\gamma}_{\; ,\sigma}
\nonumber \\
& \mbox{} & \hspace{-0.9cm}
+ (D-2) {\widetilde g}^{\alpha\beta} \, {\widetilde g}^{\rho\sigma}
  \, \kappa h_{\beta\gamma} u_{\sigma} c^{\gamma}_{\; ,\rho}
- (D-2) {\widetilde g}^{\alpha\beta} \, 
  {\widetilde g}^{\rho\sigma}_{\;\; ,\gamma}
  \, \kappa h_{\beta\rho} u_{\sigma} c^{\gamma} \big\}
\nonumber \\
& \mbox{} & \hspace{-0.9cm}
- 2(D-2) a^D {\widetilde H}^2 \sqrt{-{\widetilde g}} \, 
  {\widetilde g}^{\alpha\beta} \, {\widetilde g}^{\rho\sigma} \, 
  \kappa h_{\beta\rho} u_{\sigma} 
  {\overline c}_{\alpha} u_{\gamma} c^{\gamma}
\; . \label{Lgh2}
\end{eqnarray}

\section{The Leading Logarithm Approximation for Pure Gravity}

The LLOG methodology was extensively analyzed in 
subsection 2.2 where the relevant rule for the pure 
gravitational case was identified:
\begin{equation}
\frac{\delta S[h]}{\delta h_{\mu\nu}} \Big\vert_{LLOG} 
\equiv
\frac{\delta S[h]_{class}}{\delta h_{\mu\nu}} \Big\vert_{stoch}
\! - 
a^4 \sqrt{-{\widetilde g}} \; T[h]^{\mu\nu} \Big\vert_{ind} 
\! = 0
\; . \label{QGrule}
\end{equation}

An important simplification can be a priori made since 
it turns out that all coincidence limits that appear in
the computations required by (\ref{QGrule}) are finite
and the only one that is not does not appear.
\footnote{All propagator coincidence limits are displayed
in the Appendix (\ref{proplimA+B}-\ref{ddproplimC}). 
Of these only $i{\widetilde \Delta}_A (x;x')\big\vert_{x=x'}$ 
is not finite.}
Hence, we can take the $D=4$ limit which shall considerably
simplify the intricate tensor algebra that must be done.
\footnote{Some of the identities used in the algebraic tensor 
steps can be found in the Appendix (\ref{ID1}-\ref{ID3}).}

\subsection{The contribution from $\mathcal{L}_{inv}
+ \mathcal{\widetilde L}_{GF}$}

It turns out that it is preferable to add the 
contributions of $\mathcal{L}_{inv}$ (\ref{Linv2}) 
and $\mathcal{\widetilde L}_{GF}$ (\ref{Lgf2}) 
because some of their terms simplify against each 
other. The six terms comprising $\, \mathcal{L}_{inv} 
\!+\! \mathcal{\widetilde L}_{GF} \,$ can be grouped
into three convenient parts:
\begin{equation}
\mathcal{L}_{inv} + \mathcal{\widetilde L}_{GF}
\; \equiv \;
\mathcal{L}_{1+2+3} + \mathcal{L}_{4+5} + \mathcal{L}_6
\; , \label{Linv+Lgf}
\end{equation}
where these three convenient parts equal:
\begin{eqnarray}
\mathcal{L}_{1+2+3} 
&\!\!\! = \!\!\!&
a^2 \sqrt{-{\widetilde g}} \, {\widetilde g}^{\alpha\beta}
{\widetilde g}^{\rho\sigma} {\widetilde g}^{\gamma\delta} 
\nonumber \\
& \mbox{} & \hspace{0.3cm}
\times \,\Big\{ \!
- \tfrac14 h_{\alpha\rho, \gamma} h_{\beta\sigma, \delta} 
+ \tfrac18 h_{\alpha\beta, \gamma} h_{\rho\sigma, \delta}
+ a^2 {\widetilde H}^2 h_{\gamma\rho} u_{\sigma} 
  h_{\delta\alpha} u_{\beta} \Big\}
\; , \qquad \label{L1+2+3} \\
\mathcal{L}_{4+5} 
&\!\!\! = \!\!\!&
\sqrt{-{\widetilde g}} \, {\widetilde g}^{\alpha\beta}
{\widetilde g}^{\rho\sigma} {\widetilde g}^{\gamma\delta}
\partial_{\beta} \Big[ 
- \tfrac12 \partial_{\sigma} \big( a^2 h_{\gamma\rho} 
  h_{\delta\alpha} \big)
+ a^2 h_{\gamma\rho} h_{\delta\alpha, \sigma} \Big]
\; , \label{L4+5} \\
\mathcal{L}_6
&\!\!\! = \!\!\!&
\kappa a^3 H \sqrt{-{\widetilde g}} \, {\widetilde g}^{\alpha\beta}
{\widetilde g}^{\rho\sigma} {\widetilde g}^{\gamma\delta}
h_{\rho\sigma, \gamma} h_{\delta\alpha} h_{0\beta}
\; . \label{L6} 
\end{eqnarray}

The reduction process described in subsection 2.2 is a 
straightforward procedure that very quickly becomes very 
cumbersome. Before presenting the results, it is perhaps
instructive to see how this develops by considering one
of the terms present in the Lagrangian, for instance the 
first term in (\ref{L1+2+3}).
\\ [5pt]
{\it (i)} Taking the variation with respect to the graviton
field gives:
\begin{eqnarray}
\frac{\delta S_{(1)}}{\delta h_{\mu\nu}(x)}
&\!\!\! \equiv \!\!\!&
\frac{\delta}{\delta h_{\mu\nu}(x)}
\Big[ \int d^4 y \, (-\tfrac14) \, a^2
\sqrt{-{\widetilde g}} \, {\widetilde g}^{\alpha\beta} \,
{\widetilde g}^{\rho\sigma} \, {\widetilde g}^{\gamma\delta} \,
h_{\alpha\rho, \gamma} h_{\beta\sigma, \delta} \Big]
\label{S1eom} \\
&\!\!\! = \!\!\!&
- \frac{\kappa}{4} \, a^2 \sqrt{-{\widetilde g}} \, \Big\{ 
\tfrac12 \, {\widetilde g}^{\alpha\beta} \, {\widetilde g}^{\mu\nu} \,
  {\widetilde g}^{\rho\sigma} \, {\widetilde g}^{\gamma\delta}
-2 \, {\widetilde g}^{\mu(\alpha} \, {\widetilde g}^{\beta)\nu} \,
  {\widetilde g}^{\rho\sigma} \, {\widetilde g}^{\gamma\delta}
\nonumber \\
& \mbox{} & \hspace{-1.7cm}
- {\widetilde g}^{\alpha\beta} \, {\widetilde g}^{\rho\sigma} \,
  {\widetilde g}^{\mu(\gamma} \, {\widetilde g}^{\delta)\nu} 
  \Big\} \, h_{\alpha\rho, \gamma} h_{\beta\sigma, \delta}
+ \frac12 \frac{\partial}{\partial x^{\delta}}
  \big[ a^2 \sqrt{-{\widetilde g}} \,
  {\widetilde g}^{\alpha(\mu} \, {\widetilde g}^{\nu)\rho} \,
  {\widetilde g}^{\gamma\delta} \, h_{\alpha\rho, \gamma} \big]
\; , \qquad \label{S1var}
\end{eqnarray}
where the variations have been performed left to right 
on the sequence of metrics in (\ref{S1eom}).
\\ [5pt]
{\it (ii)} One of the operations inherent in the LLOG approximation
is {\it ``integrating out"} differentiated graviton fields, as
for instance in the first of the two terms in (\ref{S1var}):
\begin{equation}
h_{\alpha\rho, \gamma} h_{\beta\sigma, \delta}
\quad \longrightarrow \quad
\partial_{\gamma} \partial'_{\delta} \,
i \Bigl[ \mbox{}_{\alpha\rho} 
{\widetilde \Delta}_{\beta\sigma} \Bigr](x;x') \Big\vert_{x=x'}
\; , \label{S1intA}
\end{equation}
where (\ref{S1intA}) is valid for constant ${\widetilde g}_{\mu\nu}$.
\\
In the second of the two terms in (\ref{S1var}) all fields
are differentiated and we must sequentially extract a graviton 
field $h_{\beta\sigma}$ from the $\sqrt{-{\widetilde g}}$ and 
the three contravariant conformally rescaled metrics that follow:
\begin{equation}
h_{\alpha\rho, \gamma} h_{\beta\sigma}
\quad \longrightarrow \quad
\partial_{\gamma} \,
i \Bigl[ \mbox{}_{\alpha\rho} 
{\widetilde \Delta}_{\beta\sigma} \Bigr](x;x') \Big\vert_{x=x'}
\; , \label{S1intB}
\end{equation}
where (\ref{S1intB}) is valid for constant ${\widetilde g}_{\mu\nu}$.
\\ [5pt]
{\it (iii)} In view of (\ref{S1var},\ref{S1intA},\ref{S1intB}) 
we get: 
\begin{eqnarray}
\frac{\delta S^{(1)}}{\delta h_{\mu\nu}(x)}
&\!\!\! \longrightarrow \!\!\!&
- \frac{\kappa}{4} a^2 \sqrt{-{\widetilde g}} \, \Big\{ 
\tfrac12 \, {\widetilde g}^{\alpha\beta} \, {\widetilde g}^{\mu\nu} \,
  {\widetilde g}^{\rho\sigma} \, {\widetilde g}^{\gamma\delta}
-2 \, {\widetilde g}^{\mu(\alpha} \, {\widetilde g}^{\beta)\nu} \,
  {\widetilde g}^{\rho\sigma} \, {\widetilde g}^{\gamma\delta}
\nonumber \\
& \mbox{} & 
- {\widetilde g}^{\alpha\beta} \, {\widetilde g}^{\rho\sigma} \,
  {\widetilde g}^{\mu(\gamma} \, {\widetilde g}^{\delta)\nu} 
  \Big\} \, \partial_{\gamma} \partial'_{\delta} \,
  i \Bigl[ \mbox{}_{\alpha\rho} 
  {\widetilde \Delta}_{\beta\sigma} \Bigr](x;x') \Big\vert_{x=x'}
\nonumber \\
& \mbox{} & \hspace{-0.9cm}
+ \frac{\kappa}{2} \partial_{\delta} 
  \Big\{ \, a^2 \sqrt{-{\widetilde g}} \Big[
\tfrac12 \, {\widetilde g}^{\beta\sigma} \, {\widetilde g}^{\alpha(\mu} \,
  {\widetilde g}^{\nu)\rho} \, {\widetilde g}^{\gamma\delta}
- {\widetilde g}^{\beta\alpha} \, {\widetilde g}^{\sigma(\mu} \,
  {\widetilde g}^{\nu)\rho} \, {\widetilde g}^{\gamma\delta}
\nonumber \\
& \mbox{} & \hspace{-0.9cm}
- {\widetilde g}^{\alpha(\mu} \, {\widetilde g}^{\nu)\beta} \,
  {\widetilde g}^{\rho\sigma} \, {\widetilde g}^{\gamma\delta}
- {\widetilde g}^{\alpha(\mu} \, {\widetilde g}^{\nu)\rho} \,
  {\widetilde g}^{\gamma\beta} \, {\widetilde g}^{\delta\sigma} \Big]
  \partial_{\gamma} \,
  i \Bigl[ \mbox{}_{\alpha\rho} 
  {\widetilde \Delta}_{\beta\sigma} \Bigr](x;x') \Big\vert_{x=x'}
  \Big\}
\; , \qquad \label{S1int}
\end{eqnarray}
The coincidence limits of the graviton propagator in (\ref{S1int}) 
are finite: 
\footnote{These limits are the result of applying 
(\ref{dproplimA+B}-\ref{ddproplimC}) to (\ref{prop}).}
\begin{eqnarray}
\partial_{\gamma} \partial'_{\delta} \,
i \Bigl[ \mbox{}_{\alpha\beta} 
{\widetilde \Delta}_{\rho\sigma} \Bigr](x;x') \Big\vert_{x=x'}
&\!\!\!\!\!\!\!=\!\!\!\!\!&
- \frac{{\widetilde H}^4}{32\pi^2} a^2 
  {\widetilde g}_{\gamma\delta} \Big\{ 3 \big[ 
  {\overline \gamma}_{\alpha\beta} {\overline \gamma}_{\rho\sigma}
+ {\overline \gamma}_{\alpha\sigma} {\overline \gamma}_{\beta\rho}
-2 {\overline \gamma}_{\alpha\rho} {\overline \gamma}_{\beta\sigma}
\big]
\nonumber \\
& \mbox{} & \hspace{-0.5cm}
+ \big[ u_{\alpha} {\overline \gamma}_{\rho\beta} u_{\sigma}
+ u_{\alpha} {\overline \gamma}_{\rho\sigma} u_{\beta}
+ u_{\rho} {\overline \gamma}_{\alpha\beta} u_{\sigma}
+ u_{\rho} {\overline \gamma}_{\alpha\sigma} u_{\beta} \big]
\nonumber \\
& \mbox{} & \hspace{-0.5cm}
+ \big( u_{\alpha} u_{\rho} + {\overline \gamma}_{\alpha\rho} \big)
\big( u_{\beta} u_{\sigma} + {\overline \gamma}_{\beta\sigma} \big)
\Big\}
\; , \label{S1Ax=x'} \\
\partial_{\gamma} \,
i \Bigl[ \mbox{}_{\alpha\rho} 
{\widetilde \Delta}_{\beta\sigma} \Bigr](x;x') \Big\vert_{x=x'}
&\!\!\!\!\!\!\!=\!\!\!\!\!&
- \frac{{\widetilde H}^3}{8\pi^2} a \, u_{\gamma} \big[
{\overline \gamma}_{\alpha\beta} {\overline \gamma}_{\rho\sigma}
+ {\overline \gamma}_{\alpha\sigma} {\overline \gamma}_{\beta\rho}
-2 {\overline \gamma}_{\alpha\rho} {\overline \gamma}_{\beta\sigma}
\big]
\; . \qquad \label{S1Bx=x'}
\end{eqnarray}
{\it (iv)} When we insert (\ref{S1Ax=x'},\ref{S1Bx=x'}) in 
(\ref{S1int}), we instantly see the huge number of elementary 
tensor contractions that we have to face from a single term 
only. That said and seen, we proceed directly to the final
results. 
\\ [5pt]
{\bf -} {\it The LLOG equation (\ref{QGrule}) from 
$\mathcal{L}_{1+2+3}$:} \\
The {\it ``stochastic reduction"} of the full field equation
from $\mathcal{L}_{1+2+3}$ furnishes the following classical 
Langevin form: 
\begin{eqnarray}
\frac{\delta S[h]^{class}_{(1+2+3)}}{\delta h_{\mu\nu}} \Big\vert_{stoch}
&\!\!\! = \!\!\!& 
\label{S_1+2+3_stoch} \\
& \mbox{} & \hspace{-3.3cm}
a^4 \sqrt{-{\widetilde g}} \, \Big\{ 
\tfrac32 \, {\widetilde g}^{00} H \Big[
  {\widetilde g}^{\rho\mu} {\widetilde g}^{\sigma\nu}
  -\tfrac12 {\widetilde g}^{\rho\sigma} {\widetilde g}^{\mu\nu} \Big]
  \big[ {\dot h}_{\rho\sigma} - {\dot \chi}_{\rho\sigma} \big]
+ 2 {\widetilde H}^2 u^{(\mu} {\widetilde g}^{\nu)(\alpha} u^{\beta)}
  h_{\alpha\beta}
\nonumber \\
& \mbox{} & \hspace{-3.3cm}
+\, \kappa {\widetilde H}^2 \Big[
\tfrac12 \, {\widetilde g}^{\mu\nu} u^{(\alpha} 
  {\widetilde g}^{\beta)(\rho} u^{\sigma)}
- 2 u^{(\mu} {\widetilde g}^{\nu)(\alpha} 
  {\widetilde g}^{\beta)(\rho} u^{\sigma)}
- u^{(\alpha} {\widetilde g}^{\beta)(\mu} 
  {\widetilde g}^{\nu)(\rho} u^{\sigma)} \Big]
  h_{\alpha\beta} h_{\rho\sigma} \Big\}
\; , \nonumber
\end{eqnarray}
where $\chi_{\mu\nu}$ denotes the stochastic jitter.
The corresponding {\it ``integrating out"} operation
leads to the following induced stress tensor:
\begin{equation}
- a^4 \sqrt{-{\widetilde g}} \; T[h]^{\mu\nu}_{(1+2+3)} \Big\vert_{ind}
\; = \;
a^4 \sqrt{-{\widetilde g}} \, \frac{\kappa {\widetilde H}^4}{8\pi^2}
\Big[\! - \tfrac12 \, {\widetilde g}^{\mu\nu} \!+ 6 u^{\mu} u^{\nu} \Big]
\; . \label{S_1+2+3_Tmn}
\end{equation}
\\ [5pt]
{\bf -} {\it The LLOG equation (\ref{QGrule}) from 
$\mathcal{L}_{4+5}$:} \\
Respective results were derived for the Langevin equation:
\begin{eqnarray}
\frac{\delta S[h]^{class}_{(4+5)}}{\delta h_{\mu\nu}} \Big\vert_{stoch}
&\!\!\! = \!\!\!& 
\label{S_4+5_stoch} \\
& \mbox{} & \hspace{-3cm}
a^4 \sqrt{-{\widetilde g}} \, \Big[\!
- {\widetilde g}^{\mu\nu} u^{(\alpha} 
  {\widetilde g}^{\beta)(\rho} u^{\sigma)}
+ 2u^{(\mu} {\widetilde g}^{\nu)(\alpha} 
  {\widetilde g}^{\beta)(\rho} u^{\sigma)}
+ 2u^{(\mu} {\widetilde g}^{\nu)(\rho} 
  {\widetilde g}^{\sigma)(\alpha} u^{\beta)}
\nonumber \\
& \mbox{} & \hspace{-1.3cm}
+\, 2u^{(\alpha} {\widetilde g}^{\beta)(\mu} 
  {\widetilde g}^{\nu)(\rho} u^{\sigma)} \Big]
\Big\{ \tfrac32 \, {\widetilde H}^2 \kappa 
   h_{\alpha\beta} h_{\rho\sigma} 
+ \frac{2 {\widetilde H}}{N} \, \kappa h_{\alpha\beta}
  \big[ {\dot h}_{\rho\sigma} - {\dot \chi}_{\rho\sigma}
  \big] \Big\}
\; , \nonumber 
\end{eqnarray}
and the induced stress-tensor emanating from $\mathcal{L}_{4+5}$
(\ref{L4+5}):
\begin{equation}
- a^4 \sqrt{-{\widetilde g}} \; T[h]^{\mu\nu}_{(4+5)} \Big\vert_{ind}
\! = 
a^4 \sqrt{-{\widetilde g}} \, \frac{\kappa {\widetilde H}^4}{8\pi^2}
\Big[\! -8 u^{(\mu} {\widetilde g}^{\nu)\rho} u^{\sigma}
+ 4 u^{\mu} u^{\nu} u^{\rho} u^{\sigma} \Big] \kappa h_{\rho\sigma} 
\; . \label{S_4+5_Tmn}
\end{equation}
\\ [5pt]
{\bf -} {\it The LLOG equation (\ref{QGrule}) from 
$\mathcal{L}_{6}$:} \\
Finally we computed the Langevin equation:
\begin{eqnarray}
\frac{\delta S[h]^{class}_{(6)}}{\delta h_{\mu\nu}} \Big\vert_{stoch}
&\!\!\! = \!\!\!& 
\label{S_6_stoch} \\
& \mbox{} & \hspace{-2.9cm}
a^4 \sqrt{-{\widetilde g}} \, \Big\{ \!
- 3 \kappa H^2 \, {\widetilde g}^{\mu\nu} {\widetilde g}^{\alpha\beta}
  {\widetilde g}^{0\gamma} h_{0\alpha} h_{\beta\gamma}
- \kappa {\widetilde H} \delta^{(\mu}_{\;\; 0} \, {\widetilde g}^{\nu)\alpha}
  u^{\beta} {\widetilde g}^{\rho\sigma} \big[ {\dot h}_{\rho\sigma} 
  - {\dot \chi}_{\rho\sigma} \big] h_{\alpha \beta}
\nonumber \\
& \mbox{} & \hspace{-1.2cm}
-\, \kappa {\widetilde H} \Big[ u^{(\mu} {\widetilde g}^{\nu)\alpha} 
  {\widetilde g}^{\rho\sigma}
- {\widetilde g}^{\mu\nu} u^{(\rho} {\widetilde g}^{\sigma)\alpha}
  \Big] \big[ {\dot h}_{\rho\sigma} - {\dot \chi}_{\rho\sigma} \big]
  h_{\alpha 0}
\nonumber \\
& \mbox{} & \hspace{-1.2cm}
+\, \kappa {\widetilde H} \, {\widetilde g}^{\mu\nu}
  u^{(\rho} {\widetilde g}^{\sigma)\alpha} {\overline \gamma}^{\beta}_{\; 0} 
  \big[ {\dot h}_{\alpha\beta} - {\dot \chi}_{\alpha\beta} \big]
  h_{\rho\sigma}
\nonumber \\
& \mbox{} & \hspace{-1.2cm}
-\, \kappa^2 {\widetilde H} \Big[\!
- u^{\gamma} {\widetilde g}^{\alpha(\mu} {\widetilde g}^{\nu)\beta}
  {\widetilde g}^{\rho\sigma}
- {\widetilde g}^{\alpha\beta} {\widetilde g}^{\rho\sigma}
  u^{(\mu} {\widetilde g}^{\nu)\gamma}
+ {\widetilde g}^{\alpha\rho} {\widetilde g}^{\beta\sigma}
  {\widetilde g}^{\mu\nu} u^{\gamma}
\nonumber \\
& \mbox{} & \hspace{+0.2cm}
+\, {\widetilde g}^{\alpha\beta} {\widetilde g}^{\mu\nu}
  u^{(\rho} {\widetilde g}^{\sigma)\gamma} \Big]
  \big[ {\dot h}_{\rho\sigma} - {\dot \chi}_{\rho\sigma} \big]
  h_{\gamma\alpha} h_{\beta 0} \Big\}
\; , \nonumber
\end{eqnarray}
and induced stress tensor coming from the remaining term 
$\mathcal{L}_6$ (\ref{L6}):
\begin{eqnarray}
- a^4 \sqrt{-{\widetilde g}} \; T[h]^{\mu\nu}_{(6)} \Big\vert_{ind}
&\!\!\! = \!\!\!& 
\label{S_6_Tmn} \\
& \mbox{} & \hspace{-4.1cm}
a^4 \sqrt{-{\widetilde g}} \, \frac{\kappa H {\widetilde H}^3}{8\pi^2}
\Big\{ 4 u^{(\mu} {\overline \gamma}^{\nu)}_{\,\, 0}
+ \big[ 12 u^{(\mu} {\widetilde g}^{\nu)\rho}
- 8 u^{(\mu} {\overline \gamma}^{\nu)\rho} \big] \kappa h_{\rho 0}
- 4 u^{(\mu} {\widetilde g}^{\nu)\rho} {\overline \gamma}^{\sigma}_{\; 0}
  \, \kappa h_{\rho\sigma}
\nonumber \\
& \mbox{} & \hspace{-3.7cm}
+\, \big[\, 2 {\overline \gamma}^{\mu\nu} {\overline \gamma}^{\rho}_{\; 0}
- 6 {\widetilde g}^{\mu\nu} {\overline \gamma}^{\rho}_{\; 0}
- 6 {\overline \gamma}^{\rho(\mu} {\overline \gamma}^{\nu)}_{\,\, 0}
+ 12 \delta^{(\mu}_{\,\, 0} {\widetilde g}^{\nu)\rho} 
- 4 \delta^{(\mu}_{\,\, 0} {\overline \gamma}^{\nu)\rho} 
\big] u^{\sigma} \kappa h_{\rho\sigma}
\nonumber \\
& \mbox{} & \hspace{-3.7cm}
+\, \big[\! 
- 12 u^{(\mu} {\widetilde g}^{\nu)\rho} {\widetilde g}^{\sigma\alpha}
+ 4 u^{(\mu} {\widetilde g}^{\nu)\alpha} {\overline \gamma}^{\rho\sigma} 
+ 4 u^{(\mu} {\overline \gamma}^{\nu)\rho} {\widetilde g}^{\sigma\alpha}
\big] \kappa^2 h_{\alpha\rho} h_{\sigma 0}
\nonumber \\
& \mbox{} & \hspace{-3.7cm}
+\, \big[\! -12 {\widetilde g}^{\rho(\mu} {\widetilde g}^{\nu)\sigma}
+ 4 {\overline \gamma}^{\rho(\mu} {\widetilde g}^{\nu)\sigma}
+ 4 {\overline \gamma}^{\sigma(\mu} {\widetilde g}^{\nu)\rho}
+ 6 {\widetilde g}^{\mu\nu} {\overline \gamma}^{\rho\sigma}
\nonumber \\
& \mbox{} & \hspace{-3.1cm}
+\, 2 {\overline \gamma}^{\mu(\rho} {\overline \gamma}^{\sigma)\nu}
- 2 {\overline \gamma}^{\mu\nu} {\overline \gamma}^{\rho\sigma}
\big] u^{\alpha} \kappa^2 h_{\alpha\rho} h_{\sigma 0} \Big\}
\; . \nonumber
\end{eqnarray}

\subsection{The contribution from $\mathcal{L}_{gh}$}

The reduction process described in subsection 2.2 when 
applied to the ghost Lagrangian $\mathcal{L}_{gh}$
(\ref{Lgh2}) must take into account that we seek an LLOG 
effective field equation for the {\it graviton} field.
Hence all the ghost fields must be integrated out:
\begin{eqnarray}
{\overline c}_{\alpha} c^{\beta}
&\!\! \longrightarrow \!\!&
i \big[ \mbox{}_{\alpha} \Delta^{\beta} \big] (x;x')
\, = \,
{\overline \delta}^{\beta}_{\; \alpha} \,
  i{\widetilde \Delta}_A (x;x')
- u^{\beta} u_{\alpha} \, i{\widetilde \Delta}_B (x;x')
\; , \label{ghostprop1} \\
{\overline c}_{\alpha} c^{\alpha}
&\!\! \longrightarrow \!\!&
i \big[ \mbox{}_{\alpha} \Delta^{\alpha} \big] (x;x')
\, = \,
3 \, i{\widetilde \Delta}_A (x;x')
+ i{\widetilde \Delta}_B (x;x')
\; , \label{ghostprop2}
\end{eqnarray}
and we then determine the coincidence limits of 
(\ref{ghostprop1},\ref{ghostprop2}) from the Appendix
identities (\ref{proplimA+B}-\ref{ddproplimC})

The only contribution to the LLOG approximation (\ref{QGrule})
coming from $\mathcal{L}_{gh}$ is the induced stress tensor
from integrating out all ghost fields using  
(\ref{ghostprop1},\ref{ghostprop2}):
\begin{eqnarray}
\frac{\delta S[h]^{class}_{gh}}{\delta h_{\mu\nu}} \Big\vert_{stoch}
&\!\!\!\!\! = \!\!\!\!&
0 
\label{S_gh_stoch} \\
- a^4 \sqrt{-{\widetilde g}} \; T[h]^{\mu\nu}_{gh} \Big\vert_{ind}
&\!\!\!\!\! = \!\!\!\!&
\kappa \, \frac{{\widetilde H}^4}{8\pi^2} \, a^4 \sqrt{-{\widetilde g}} 
\, \Big\{ \big[ 
- \tfrac32 + {\overline \gamma}^{\rho\sigma} \kappa h_{\rho\sigma}
+ u^{\rho} u^{\sigma} \kappa h_{\rho\sigma} 
\big] \, {\widetilde g}^{\mu\nu}
\nonumber \\
& \mbox{} & \hspace{-0.5cm}
+ \big[ -4 + 2 {\overline \gamma}^{\rho\sigma} 
\kappa h_{\rho\sigma} \big] \, u^{\mu} u^{\nu}
\nonumber \\
& \mbox{} & \hspace{-0.5cm}
+ \big[ -2 {\overline \gamma}^{\mu\rho} \, {\overline \gamma}^{\nu\sigma}
- 5 u^{(\mu} \, {\overline \gamma}^{\nu)\rho} u^{\sigma}
+ 7 u^{\mu} u^{\nu} u^{\rho} u^{\sigma} \big] \, \kappa h_{\rho\sigma}
\Big\} 
\; . \qquad \label{S_gh_Tmn}
\end{eqnarray}

\section{Epilogue}

The continual production of inflationary gravitons tends 
to endow graviton loop corrections with secular growth 
factors. During a prolonged period of inflation these 
factors eventually overwhelm the small loop-counting 
parameter of $G H^2$, which causes perturbation theory 
to break down. Developing a re-summation technique that 
permits one to evolve to late times has been a long 
struggle owing to the derivative interactions of gravity. 
The analogous problem for non-linear sigma models was 
recently solved by combining a variant of Starobinsky's 
stochastic formalism with a variant of the renormalization 
group \cite{Miao:2021gic}. The technique has recently been 
generalized to scalar corrections to gravity \cite{Miao:2024nsz}, 
and here we extended it to pure gravity.

The basic assumption in our analysis is that changes in the
geometrical background are significantly slower than the changes
in the scale factor. The basic result is the derivation of
the Langevin equation for pure quantum gravity on de Sitter
background. The basic advantage of this equation is that the
leading logarithm (LLOG) approximation it represents is valid
to {\it all} orders of perturbation theory.

To achieve this entailed doing three things: \\
{\it (i)} Generalizing the gauge fixing condition by replacing
$\eta_{\mu\nu}$ with $\widetilde{g}_{\mu\nu}$. \\
{\it (ii)} Integrating out differentiated graviton fields to
produce a leading logarithm stress tensor.
\footnote{The key observation here is that constant
$\widetilde{g}{\mu\nu}$ corresponds to a de Sitter geometry
with $H^2 \rightarrow \widetilde{H}^2$.} \\
{\it (iii)} Stochastically simplifying the classical equation.

We must postpone the variant of the renormalization group 
analysis for pure gravity because we must first determine 
the loop-dependent counterterms in the new gauge (\ref{F}), 
but we have been able to derive herein the Langevin equation 
needed to implement the stochastic part of the re-summation 
technique.

Although correct, the equations we have derived require 
further processing in order to delete differentiated fields 
from equations which contain the same field undifferentiated. 
We must untangle the constrained $h_{\mu 0}$ components from 
the stochastically fluctuating $h_{ij}$ components. Fluctuations 
in the latter drive evolution in the former, and this evolution 
changes the strength of stochastic fluctuations. These changes 
will simplify the equations considerably and clarify their 
physical effect, however, implementing them is a considerable 
undertaking which we defer to a later work. 

After simplifying the LLOG equations we derived by focusing
on the interaction of the fluctuating with the constrained
metric tensor degrees of freedom, we can use them to solve
quite a few problems. For instance, the standard perturbative 
expansion can be easily recovered from the LLOG equations 
and our re-summation prescription can be tested for its 
correspondence limit with directly obtained perturbative 
results: \\
{\bf -} We must re-compute the 1-loop graviton 1PI 1-point 
function \cite{Tsamis:2005je} in the new gauge because this 
will provide a very stringent check on our induced stress 
tensor. \\
{\bf -} We also need to re-compute the 1-loop graviton 1PI 
2-point function \cite{Tan:2021ibs} in the new gauge. This 
will determine the gauge-dependent coefficients of the $R^2$ 
and $C^2$ counterterms. \\
{\bf -} Our constant $\widetilde{g}_{\mu\nu}$ propagator could 
be used to evaluate non-coincident diagrams in the leading 
logarithm approximation. \\
In general, after the perturbative correspondence limit of our 
Langevin equation has been thoroughly tested, the equation can 
be used to compute in a much simpler way the leading logarithm 
contribution of diagrams beyond 1-loop order.

By far the most interesting physical applications are those 
where perturbative results exhibit a secular effect which
leads to a breakdown of perturbation theory. In that case,
the LLOG approximation may be proven to be precisely what
is needed to rectify the breakdown by re-summing all orders
of perturbation theory. To solve the Langevin equation of 
the stochastic field, we note that expectation values of
functionals of the stochastic field can be computed in terms 
of a probability density which satisfies a Fokker-Planck 
equation derived from the Langevin equation of the stochastic
field \cite{Tsamis:2005hd}. The Fokker-Planck equation is 
usually too hard to solve but this may not be the case 
provided that the asymptotic behaviour of the probability 
density is appropriate. \\
{\bf -} For the case of the scalar theory (\ref{Lscalar}) 
with $V(\phi) \sim \lambda \phi^4$, the assumption that 
the probability distribution $\varrho(t, \varphi)$ 
asymptotically reaches a constant $\varrho_{\infty}(\varphi)$
does provide a solution: the evolution stops, we reach 
a de Sitter spacetime with a slightly higher $\Lambda$ 
and, we remind again, this is an all-orders result 
exhibiting the full behaviour of the particular scalar 
theory at late times \cite{Starobinsky:1994bd,
Tsamis:2005hd}. \\
{\bf -} For the case of pure quantum gravity, we expect 
\cite{Tsamis:1994ca,Tsamis:2011ep} the assumption that 
asymptotically we reach a constant does {\it not} provide 
a solution and, therefore, time evolution never stops.

More specifically, we expect that the Hubble parameter 
$H(t)$ observable which initially was a constant diminishes
with time due to a secular effect whose ultimate physical
origin is the universally negative nature of gravitation;
in this case the negative interaction energy among the 
copious gravitons produced during inflation. The gauge 
fixing dependence of all results is obviously a serious
physical issue that must be addressed and resolved. It is 
worth noting that in all cases analyzed so far the leading
infrared effect is {\it not} a gauge artifact 
\cite{Miao:2017feh,Glavan:2024elz}.

A strong indication that inflation is eventually extinguished 
can emerge from studying the stochastically driven behavior 
of $h_{00}$. If one initially sets the Hubble parameter $H$ 
to its constant, de Sitter value and the equations we have 
derived imply that $h_{00}$ grows to the point that 
$\, \kappa h_{00} \rightarrow 1 - \varepsilon$, then we have 
$\, \widetilde{g}_{00} \rightarrow -1 + (1 - \varepsilon) = 
-\varepsilon$. This makes the factors of $\widetilde{g}^{00}$ 
in $\widetilde{H}$ grow very large. The physical interpretation 
is that the universe is almost within its Schwarzschild radius, 
poised on the verge of gravitational collapse. At this point 
accelerated expansion must stop, and there are no more 
stochastic fluctuations to drive further evolution. 

Moreover, it seems perfectly possible to further generalize 
the formalism to cover this post-inflationary epoch. This can 
been accomplished for the case of non-linear sigma models that 
are spectators to a scalar-driven inflation which merges into 
a $\Lambda$CDM universe \cite{Kasdagli:2023nzj,Woodard:2023cqi}.

Nonetheless, one must entertain the possibility that the 
LLOG approximation - although non-perturbative - still 
cannot demonstrate the screening effect or capture its 
time evolution. Thus, another more appropriate technique 
is needed {\it or} the screening effect is simply absent. 
We shall find out.

\vskip 0.5cm

\centerline{\bf Acknowledgements}

This work was partially supported by Taiwan NSTC grants 
111-2112-M-006-038 and 112-2112-M-006-017, by NSF grant 
PHY-2207514 and by the Institute for Fundamental Theory 
at the University of Florida.

\vskip 1cm

\section{Appendix: Useful Identities}

\vskip 0.5cm

{\bf *} Some relations from the 3+1 decomposition:
\begin{eqnarray}
& \mbox{} &
{\widetilde H} = \frac{H}{N}
\quad , \quad 
{\widetilde H}^2 = \frac{H^2}{N^2} 
= - {\widetilde g}^{00} H^2 
\; , \label{Htilde} \\
& \mbox{} &
\delta^0_{\, \mu} = - \frac{1}{N} u_{\mu}
\quad , \quad
{\widetilde g}^{0\mu} = - \frac{1}{N} u^{\mu}
\quad , \quad
{\widetilde g}^{\mu\nu} u_{\mu} u_{\nu} 
= -1 =
{\widetilde g}_{\mu\nu} u^{\mu} u^{\nu} 
\; . \qquad\qquad \label{u's}
\end{eqnarray}
\\ [3pt]
{\bf *} Some tensor algebra identities:
\begin{equation}
{\widetilde g}^{\mu\nu} =
\eta^{\mu\nu} - \kappa h^{\mu}_{\; \alpha} \; {\widetilde g}^{\alpha\nu}
\; , \label{ID1}
\end{equation}
\begin{equation}
\eta_{\mu\nu} =
{\overline \eta}_{\mu\nu} - \delta^0_{\; \mu} \delta^0_{\; \nu}
\quad , \quad
\delta^{\mu}_{\; \nu} =
{\overline \delta}^{\mu}_{\nu} + \delta^{\mu}_{\; 0} \delta^0_{\; \nu}
\; , \label{ID2}
\end{equation}
\begin{equation}
\delta^{\mu}_{\; \alpha} \delta^{\nu}_{\; \beta}
=
{\overline \delta}^{\mu}_{\; (\alpha} 
{\overline \delta}^{\nu}_{\; \beta)}
+ 2 \delta^{(\mu}_{\;\; 0} \, {\overline \delta}^{\nu)}_{\; (\alpha}
  \delta^0_{\; \beta)}
+ \delta^{\mu}_{\; 0} \delta^{\nu}_{\; 0}
  \delta^0_{\; \alpha} \delta^0_{\; \beta}
\; . \label{ID3}
\end{equation}
\\ [3pt]
{\bf *} The various propagator coincident limit identities:
\begin{eqnarray}
& \mbox{} &
i{\widetilde \Delta}_A (x;x')\big\vert_{x=x'}
= \frac{{\widetilde H}^2}{4\pi^2} \ln a + ``\infty"
\quad , \quad
i{\widetilde \Delta}_B (x;x')\big\vert_{x=x'}
= -\frac{{\widetilde H}^2}{16\pi^2}
\; , \qquad\quad \label{proplimA+B} \\
& \mbox{} &
i{\widetilde \Delta}_C (x;x')\big\vert_{x=x'}
= +\frac{{\widetilde H}^2}{16\pi^2}
\; , \label{proplimC} \\
& \mbox{} &
\partial'_{\sigma} \, i{\widetilde \Delta}_A (x;x')\big\vert_{x=x'}
= - \frac{{\widetilde H}^3}{8\pi^2} \, a u_{\sigma}
\quad , \quad
\partial'_{\sigma} \, i{\widetilde \Delta}_B (x;x')\big\vert_{x=x'}
= 0
\; , \label{dproplimA+B} \\
& \mbox{} &
\partial'_{\sigma} \, i{\widetilde \Delta}_C (x;x')\big\vert_{x=x'}
= 0
\; , \label{dproplimC} \\
& \mbox{} &
\partial_{\rho} \partial'_{\sigma} \,
i{\widetilde \Delta}_A (x;x')\big\vert_{x=x'}
= - \frac{3 {\widetilde H}^4}{32\pi^2} a^2 \, {\widetilde g}_{\rho\sigma}
\; , \label{ddproplimA} \\
& \mbox{} &
\partial_{\rho} \partial'_{\sigma} \,
i{\widetilde \Delta}_B (x;x')\big\vert_{x=x'}
= \frac{{\widetilde H}^4}{32\pi^2} a^2 \, {\widetilde g}_{\rho\sigma}
\; , \label{ddproplimB} \\
& \mbox{} &
\partial_{\rho} \partial'_{\sigma} \,
i{\widetilde \Delta}_C (x;x')\big\vert_{x=x'}
= - \frac{{\widetilde H}^4}{32\pi^2} a^2 \, {\widetilde g}_{\rho\sigma}
\; . \label{ddproplimC}
\end{eqnarray}
{\it Note:} In the above relations, we first take 
the derivatives and then the coincidence limit 
$x' \rightarrow x$. 
\\ [19pt]
{\bf *} The d' Alembertian operator equals:
\begin{equation}
{\widetilde D}_A 
\equiv
\partial_{\alpha} \big[ a^{D-2} {\sqrt {-\widetilde g}} \,
{\widetilde g}^{\alpha\beta} \partial_{\beta} \big]
\; , \label{D_A}
\end{equation}
and its operation on the three kinds of scalar propagators
gives:
\begin{eqnarray}
{\widetilde D}_A \, i {\widetilde \Delta}_A(x;x') 
&\!\!\!=\!\!\!& 
i \delta^D(x-x')
\; , \label{iDeltaA} \\
{\widetilde D}_A \, i {\widetilde \Delta}_B(x;x')
&\!\!\!=\!\!\!& 
i \delta^D(x-x') 
+ (D-2) {\widetilde H}^2 a^D {\sqrt {-\widetilde g}} \;
  i {\widetilde \Delta}_B(x;x')
\; , \label{iDeltaB} \\
{\widetilde D}_A \, i {\widetilde \Delta}_C(x;x')
&\!\!\!=\!\!\!& 
i \delta^D(x-x') 
+ 2(D-3) {\widetilde H}^2 a^D {\sqrt {-\widetilde g}} \; 
  i {\widetilde \Delta}_C(x;x') 
\; . \qquad \label{iDeltaC}
\end{eqnarray}

\end{document}